# Phonon Lifetimes and Thermal Conductivity of the Molecular Crystal α-RDX[1]

Gaurav Kumar, Francis G. VanGessel, Daniel C. Elton & Peter W. Chung

*Department of Mechanical Engineering, University of Maryland, College Park, Maryland USA*

ABSTRACT

*The heat transfer properties of the organic molecular crystal α-RDX were studied using three phonon-based thermal conductivity models. It was found that the widely used Peierls-Boltzmann model for thermal transport in crystalline materials breaks down for α-RDX. We show this breakdown is due to a large degree of anharmonicity that leads to a dominance of diffusive-like carriers. Despite being developed for disordered systems, the Allen-Feldman theory for thermal conductivity actually gives the best description of thermal transport. This is likely because diffusive carriers contribute to over 95% of the thermal conductivity in α-RDX. The dominance of diffusive carriers is larger than previously observed in other fully ordered crystalline systems. These results indicate than van-der Waals bonded organic crystalline solids conduct heat in a manner more akin to amorphous materials than simple atomic crystals.*

## INTRODUCTION

Forming a more complete understanding of phonon dynamics in complex molecular crystals, such as the energetic α-RDX, is of fundamental importance due to the role that phonons play in initiation mechanisms of energetics [1, 2]. In molecular crystals, commonly used in energetics, phonons facilitate the formation of the microscale reaction zones, i.e. hotspots, wherein the flow of energy into key vibrational modes initiates the chemical decomposition process [1, 3, 4]. However, open questions remain as to the exact mechanism through which inter-modal and spatial energy transfer occurs in molecular crystals. Dlott et al. postulated a multiphonon up-pumping mechanism to explain the inter-modal energy transfer phenomenon, while other works speculated that the energy transfer process occurs through a direct route [1, 2]. Furthermore, there exists a relative lack of knowledge regarding how thermal energy is conducted in van der-Waals bonded organic molecular crystals in relation to "simple" atomic crystalline materials such as Si or Ge. Highlighting the possible shortcomings of the existing kinetic theory for thermal transport in molecular crystals is the extremely low thermal conductivity of α-RDX, TATB, and β-HMX, ≤ 1 W/m·K, [5, 6]. Such values are more akin to conductivities observed in amorphous or glassy materials than in atomic crystalline systems. A similar phenomenon

---

of ultralow thermal conductivity has been observed in certain inorganic perovskite and selenide compound crystals [7, 8]. Theoretical and experimental studies of both of these systems have shown a breakdown of Peierls-Boltzmann theory [8, 9, 10]. Peierls-Boltzmann theory is obtained from a partial Hamiltonian and breaks down when the full crystal Hamiltonian is sufficiently anharmonic [11]. This theory represents the prevailing approach for calculation of bulk thermal conductivity in crystalline solids, often calculated through the so-called phonon gas model (PGM). The breakdown of the PGM in the perovskite and selenide compounds has been linked to the strong anharmonicity present in those systems. Molecular crystals, notably RDX, possess similarly large unit cells that are generally correlated with a highly anharmonic Hamiltonian. Therefore, the PGM, as well as alternative thermal conductivity models, must be evaluated for their accuracy in modelling phonon mechanisms in molecular crystals. Only then will it be possible to elucidate the manner in which phonons store, transport, and transfer energy in complex molecular crystalline systems.

In this paper we evaluate the strength of anharmonic coupling in the molecular crystal $\alpha$-RDX via the phonon linewidths and lifetimes. Subsequently, using harmonic and anharmonic phonon properties, we evaluate three thermal conductivity models, namely the PGM, Cahill Watson and Pohl (CWP) model, and Allen and Feldman (AF) model. We assess the accuracy of these models in estimating the thermal conductivity of $\alpha$-RDX, comparing to values obtained from Green-Kubo molecular dynamics (GK-MD) [5]. The PGM has been shown to give excellent predictions of bulk thermal conductivity in a wide range of simple atomic crystalline systems [12, 13]. In contrast, the CWP model was initially developed for application to weakly disordered (i.e. mixed-species) crystals [14], but has shown to give better predictions than the PGM model for certain Se compounds [8]. Finally, the theory of Allen and Feldman [15] was initially developed for disordered phases in which heat transfer is diffusive in nature. However, it has been shown that significant heat currents may be carried by diffusive modes in certain "complex" crystalline materials [9].

In the next section, we present the methodology for calculating phonon linewidths, lifetimes, frequencies, and group velocities. We then apply this methodology, analysing the phonon lifetimes of $\alpha$-RDX and calculating the thermal conductivity predictions of the PGM, CWP, and AF models respectively. In the final section we present our conclusions on the presented results and discuss what they reveal about the manner in which thermal energy is transported.

**METHODOLOGY**

**Phonon relaxation times**

Accurate prediction of properties like thermal conductivity require knowledge of anharmonic vibrational properties of the material, in particular phonon lifetimes. Furthermore, the degree of anharmonicity present in a crystalline system may be used to determine the validity of the PGM [8, 9]. In this study we use the normal mode decomposition (NMD) method to determine the phonon lifetimes for all branches. The NMD approach is computationally more efficient than other techniques [16] and has been used successfully for many ordered materials such as silicon [17] and carbon nanotubes [18]. Here we give a brief overview of the methodology, while further details can be found in our earlier work [19]. Following the procedure laid out by Larkin [16] we calculate the mode projected phonon spectral energy density (SED) as

$$\Phi(\mathbf{k}, \omega) = \left| \frac{1}{\sqrt{2\pi}} \int_{-\infty}^{\infty} \dot{q}_\phi \exp(-i\omega t)\, dt \right|^2. \quad (1)$$

Here $\dot{q}_\phi(t)$ is time derivative of phonon normal mode coordinate, $\phi$ is index for phonon mode and ω is phonon mode frequency. The SED for each mode contains a single peak which was fit to a Lorentzian function using the ansatz [19]

$$\Phi(\mathbf{k}, \omega) = \sum_\lambda^{3n} C_{o\phi} \frac{\frac{\Gamma_\phi}{\pi}}{(\omega_\phi - \omega)2 + \Gamma_\phi^2}. \quad (2)$$

Phonon relaxation time can be extracted from the SED linewidth, $\Gamma_\phi$, using the relation $\tau_\phi = 1/(2\Gamma_\phi)$.

### Harmonic phonon properties

In addition to the anharmonic properties discussed above, several harmonic phonon properties are required. These are the phonon frequencies, $\omega_\phi$, group velocities, $\mathbf{v}_\phi$, and specific heat, $C_\phi$. The phonon frequencies are obtained using the harmonic lattice dynamics approach implemented in the open source package GULP [20]. The group velocities are calculated from the frequencies using a central difference scheme. The specific heat is determined from the relation

$$C_\phi = \frac{\hbar^2 \omega_\phi^2}{k_B T^2} \frac{\exp\left(\frac{\hbar \omega_\phi}{k_B T}\right)}{\left[\exp\left(\frac{\hbar \omega_\phi}{k_B T}\right) - 1\right]^2} \quad (3)$$

where $\hbar$ is reduced Planck's constant, $k_B$ is Boltzmann constant, and $T$ is temperature. From the group velocity and the phonon lifetime, the phonon mean free path (MFP) can be calculated using $\Lambda_\phi = v_\phi \tau_\phi$, where $v_\phi = |\mathbf{v}_\phi|$. All phonon properties are calculated using the Smith and Bharadwaj flexible molecule potential [21] on a uniform grid of wavevectors in the first Brillouin zone (BZ). Further information regarding calculation of these properties may be found in [22]. We have not examined the role played by the choice of the Smith and Bharadwaj potential but we note that the final estimates of thermal conductivity agree qualitatively with independently measured values. We expect, however, that better accuracy will generally be attainable using first principles or electronic structure methods.

### RESULTS & DISCUSSION

### Phonon anharmonicity in $\alpha$-RDX

In Figure 1, the calculated phonon lifetimes are shown for each vibrational mode of $\alpha$-RDX (except 3 acoustic modes at gamma point) against frequency. Our results reveal that lifetimes in $\alpha$-RDX decrease with increasing frequency up to ~3 THz, followed by flattening and subsequent increasing behaviour in lifetimes with respect to frequency. In fact, some optical modes have lifetimes comparable to or even greater than lifetimes of acoustic modes. In contrast, the acoustic modes of simple crystals such as silicon [23], carbon nanotubes [24] have the largest lifetimes while the optical modes have significantly smaller lifetimes. In fact, Callaway's model [25] for phonon lifetimes is an inverse relationship with $\omega$.

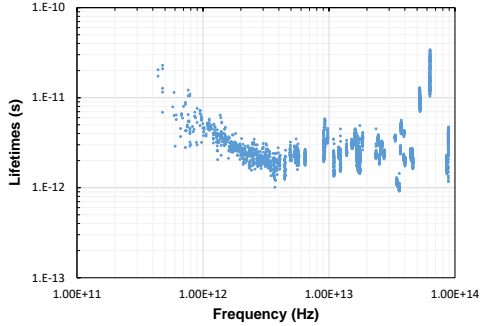

Figure 1 Phonon mode lifetimes of 2x2x2 supercell of α-RDX

The larger lifetimes of some optical modes are likely due to the large band gaps that exist in the band structure. Due to these large gaps, high frequency optical modes (>20 THz) participate in relatively fewer three- and four- phonon scattering events than lower frequency optical, and acoustic modes. Similarly, a decrease in the number of allowed phonon-phonon interactions due to an increase in the acoustic-optical bandgap is observed in III-V materials [26] Calculations to verify these results using exact scattering models are currently underway which we intend to publish in a forthcoming paper.

Note also that the magnitudes of the phonon lifetimes are quite low, on the order of 10 picoseconds, closer to lifetimes in amorphous solids than atomic crystals that have phonon lifetimes on the order of nanoseconds [23]. The short lifetimes indicate that RDX is a strongly anharmonic crystal.

## Phonon gas model

The thermal conductivity based on PGM is obtained from a sum over the contributions of all phonon carriers, i.e.

$$\kappa_{PGM} = \sum_{\phi} C_\phi v_\phi \Lambda_\phi \qquad (4)$$

For an accurate calculation of the thermal conductivity we use a uniform sampling of $15 \times 15 \times 15$ wavevectors within the first Brillouin zone for calculation of the harmonic properties. The phonon lifetimes are assumed to be constant within each band and are taken from their Gamma point value. Table I shows the predicted $\kappa_{PGM}$ of $\alpha$-RDX along 3 crystallographic directions $a$, $b$, and $c$ using PGM, along with the GK-MD value for comparison based on the same molecular potential [5]. The GK-MD approach accounts for all orders of anharmonicity and can accurately predict $\kappa$ even for condensed phases in which the PGM break down. It is clear from Table I that PGM under predicts the thermal conductivity of $\alpha$-RDX by an order of magnitude. To understand the cause of this discrepancy we appeal to the underlying assumptions of the PGM. The PGM assumes that a phonon wavepacket interacts only weakly with other phonons and therefore propagates through many unit cells before experiencing a scattering event. Quantitatively, this corresponds to the requirement that the phonon MFP be much larger than the lattice constant, i.e. the phonon wavepacket must "sample" the periodicity of the lattice [27]. In Figure 2 we plot the phonon MFP with respect to frequency and compare this value to the average lattice constant in RDX. We clearly see that except for a relatively small number (< 1%) of low-frequency phonon modes, the vast majority of carriers have a MFP smaller than the lattice constant, i.e. these carriers fall within the Ioffe-Regel regime [28]. Thus for these *diffusive* carriers the PGM is *not* a valid descriptor for how they transport thermal energy and therefore under-predicts their contribution to thermal conductivity. We note that in [19] analysis of propagating and diffusive carrier contributions to thermal conductivity were calculated within the PGM framework. However, in that study the phonon lifetime was treated as a fitting parameter in order to match the GK-MD values of $\kappa$. It is clear that an accurate thermal conductivity model for molecular crystals should account for the highly anharmonic, diffusive thermal carriers present in these systems. We now evaluate one such model, namely the CWP model for minimum thermal conductivity.

Table I. Thermal conductivity predictions of α-RDX from Green-Kubo MD and PGM, along the three principal crystallographic direction (values in units of W/m·K).

|  | a | b | c |
|---|---|---|---|
| $\kappa_{PGM}$ | 0.014 | 0.011 | 0.023 |
| $\kappa_{GK-MD}$ | 0.387 | 0.353 | 0.394 |

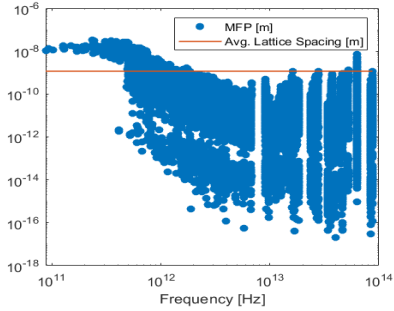

Figure 2 Comparison of phonon MFP to average lattice spacing.

### **Cahill-Watson-Pohl model**

The CWP formula was initially developed to describe the low thermal conductivity observed in crystalline alloys where the lattice structure remains intact but there is randomness in regards to the species lying at each lattice site. It can account for the dispersive nature of the acoustic mode carriers. Recently, the CWP model has been successful in predicting a qualitatively accurate thermal conductivity versus temperature trend in a Se compound with zero disorder. The improvement over PGM was due to the treatment of carriers that fall below the Ioffe-Regel limit with the CWP model, while propagating carrier contributions were still calculated using the PGM. The CWP model for thermal conductivity is [14]

$$\kappa_{CWP} = \left(\frac{\pi}{6}\right)^{\frac{1}{3}} k_B n^{\frac{2}{3}} \sum_\lambda c_\lambda \left(\frac{T}{\theta_\lambda}\right)^2 \int_0^{\frac{\theta_\lambda}{T}} \frac{x^3 e^x}{(e^x - 1)^2} dx . \qquad (5)$$

where the sum is restricted to the 3 acoustic phonon branches, indexed by $\lambda$, $c_\lambda$ is speed of sound, $n$ is the number density of atoms, and $\theta_\lambda = c_\lambda \left(\frac{\hbar}{k_B}\right)(6\pi^2 n)^{\frac{1}{3}}$. Applying the CWP formula to αRDX yields a conductivity estimate of 0.167 W/m-K. This is still 50% of the averaged $\kappa$ predicted using GK-MD which is ~0.378 W/m-K [5]. Although CWP performs markedly better than PGM, the disagreement with GK-MD is still relatively large. The cause is that the CWP model only accounts for the dispersion contributions in the three acoustic branches. However, due to the large number of optical modes, and small interband spacing, a significant fraction of heat is carried by the optical branches through coherence or non-diagonal effects [9, 11, 27]. This leads us to apply the AF model in order to account for such contributions to thermal transport in α-RDX.

**Allen-Feldman model**

Due to the majority of $\alpha$-RDX carriers falling in the Ioffe-Regel regime, the phonons are likely carrying thermal energy in a manner more akin to amorphous solids than atomic crystals. To test this, we applied the theory of Allen and Feldman (AF) in which heat is assumed to move by coherences, i.e. interband effects. In this method $\kappa$ is calculated from [15]

$$\kappa_{AF} = \frac{1}{V} \sum_{\phi} C_\phi D_\phi \quad (6)$$

Here $D_\phi$ is mode diffusivity defined as

$$D_\phi = \frac{\pi V^2}{3\hbar^2 \omega_\phi^2} \sum_{\phi'}^{\phi' \neq \phi} |S_{\phi\phi'}|^2 \delta(\omega_\phi - \omega_{\phi'}) \quad (7)$$

where $S_{\phi\phi'}$ is the off diagonal term of the heat current operator [11] and $\delta$ is the Dirac delta function that enforces energy conservation. The diffusivity is an intrinsic property of the normal modes and requires no assumption about the propagating nature of the phonons, such as in the PGM. We calculate the phonon mode diffusivity using a $3 \times 3 \times 3$ supercell in GULP [20]. The size of the supercell was chosen to balance the computational cost associated with the simulation and larger simulations are currently underway. Thermal conductivity from the AF approach is 0.354 W/m-K which is within 6% of the GK-MD estimate [5], see Table II below. The marked improvement of the AF prediction of $\kappa$ relative to PGM and CWP indicates strongly that diffusive-like carriers contribute to the majority of thermal transport in $\alpha$-RDX.

Table II Comparison of κ predicted by various thermal conductivity models (unit W/mK)

| $\kappa_{GK-MD}$ | $\kappa_{AF}$ | $\kappa_{CWP}$ | $\kappa_{PGM}$ | $\kappa_{AF} + \kappa_{PGM}$ |
|---|---|---|---|---|
| 0.378 | 0.354 | 0.167 | 0.017 | 0.371 |

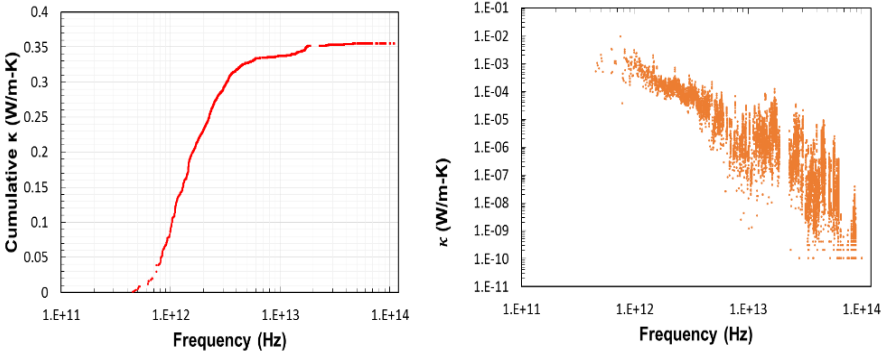

Figure 3. Cumulative and modal contributions to thermal conductivity with respect to frequency in α-RDX. Values calculated using Allen-Feldman theory.

Within the AF theory, we can break down the contributions to $\kappa_{AF}$ by individual carrier's accumulation and by each carrier separately. These are shown respectively in Figure 3. We observe that the majority of heat is carried by low frequency phonon modes and the contribution from high frequency modes is negligible. According to Figure 3, 11% of the low frequency modes account for 90% of the total $\kappa$. This is a consequence of the fact that high frequency optical phonons have very small specific heat ($C_\phi \to 0$ when $\hbar\omega_\phi \gg k_B T$) which combined with their relatively small diffusivity results in very small contribution to $\kappa$. It is also interesting to note that the AF theory considers heat transport purely due to the off-diagonal terms of heat current operator, which are non-propagating, while PGM considers heat transport via propagating phonon carriers (i.e. diagonal component). Within the unified approach [9], the propagating and non-propagating contributions are additive (termed the Peierls and coherence contributions respectively in that work). Thus, if we combine contributions from both $\kappa_{AF}$ and $\kappa_{PGM}$, we obtain a thermal conductivity of 0.371 W/m·K which is within 2% the GK-MD prediction. Note that this comparison may suffer from insufficient BZ sampling as a $3 \times 3 \times 3$ supercell was used for AF approach while using $15 \times 15 \times 15$ grid of wavevectors was used for calculation of the propagating contribution. However, it is evident that both propagating and diffusive carriers contribute substantially to thermal transport in $\alpha$-RDX, where the diffusive component is dominant, contributing 95% to the total thermal conductivity.

## CONCLUSION

In this work we have presented our preliminary findings on the nature of thermal transport in the complex molecular crystal $\alpha$-RDX. Analysis of the phonon linewidths and lifetimes indicate that $\alpha$-RDX is a highly anharmonic crystal. The inability of the PGM to accurately estimate thermal conductivity indicates that the strong anharmonicity leads to a breakdown in the Peierls picture for thermal transport. Table II summarizes thermal conductivity calculated using the various approaches used in this study. For $\alpha$-RDX, Allen-Feldman harmonic theory performs best. This result is intuitive as AF theory involves fewer assumption than PGM or CWP. We observed that low frequency optical phonons play a significant role in carrying heat in $\alpha$-RDX. It is notable that these results suggest that diffusive carriers are the primary mechanism of heat transport in $\alpha$-RDX, where propagating phonon modes contribute less than 5%. Similar phenomena involving a partitioning of thermal transport between diffusive and propagating modes has been previously observed in inorganic crystalline materials. However the relative contribution of diffusive carriers in $\alpha$-RDX is significantly larger than previously observed in other crystalline systems. We suspect these observation regarding the nature of heat transfer in $\alpha$-RDX will extend to other organic molecular crystals, particularly those with applications to energetics technology. In such systems, accurate description of thermal transport requires accounting for both acoustic and optical phonon bands, as well as both propagating and diffusive phonon modes.

## ACKNOWLEDGMENTS

G. K. and F. V. gratefully acknowledge the graduate fellowship from the Center for Engineering Concepts Development. This work was also supported, in part, by the Army Research Office under Award W911NF-14-1-0330 and the Department of Mechanical Engineering at the University of Maryland.